%% file: PZ.tex
\def\IDvalue{PZ}
\def\DIRvalue{PestunZabzine}
\def\titlevalue{Introduction to localization in quantum field theory}
\def\authorvalue{Vasily Pestun$^1$ and Maxim Zabzine$^2$}
\def\shortauthorvalue{Vasily Pestun and Maxim Zabzine}
\def\addressvalue{$^1$Institut des Hautes \'Etudes Scientifique, France\\
$^2$Department of Physics and Astronomy, Uppsala University, Sweden\\
\tt pestun@ihes.fr, Maxim.Zabzine@physics.uu.se }
\def\abstractvalue{ This is the introductory chapter to the volume. 
We review the main idea of the localization technique and its brief history both in geometry and in QFT. 
 We discuss localization in diverse dimensions and give an overview of the major applications of 
  the localization calculations for  supersymmetric theories.  
 We explain the focus of the present volume.
}
\def\preprintvalue{}

\ifx\ifLONG\undefined
\documentclass[12pt,psamsfonts,reqno]{amsart} 
\input{header.tex}

\input{PEdef.tex}
\begin{document}
\documentheader
\else 
\chapterheader 
\fi

\section{Main idea and history}

According to the English
dictionary\footnote{E.g. \url{http://www.dictionary.com}, 
  based on the Random House Dictionary, \copyright Random House, Inc. 2016} the word \emph{localize} means to make
local, fix in or assign or restrict to a particular place,
locality. Both in mathematics and physics the word ``localize" has
multiple meanings and typically  physicists with different
backgrounds mean different things by  localization.  This volume
is devoted to the extension of the Atiyah-Bott localization formula (and
related statements, e.g.  the Duistermaat-Heckman formula and different versions of the fixed-point theorem) in differential geometry
to an infinite dimensional situation of path integral,  and in particular in the context of
supersymmetric quantum field theory.  In quantum field theory one says
``supersymmetric localization" to denote such computations.  In this volume we concentrate on the development of
the supersymmetric localization technique during the last ten years, 2007-2016.

 In differential geometry the idea of localization can be traced back to
 1926 \cite{PZMR1501331}, when Lefschetz proved the fixed-point formula
 which  counts fixed points of a continuous map of a topological space
 to itself by using the graded trace of the induced map on  the homology
 groups of this space.   In the 1950's, the Grothendieck-Hirzebruch-Riemann-Roch theorem
expressed  in the most general form  the index of a holomorphic
vector bundle (supertrace over graded cohomology space) in terms of certain characteristic classes. 
In the 1960's, the Atiyah-Singer index theorem solved the same problem
for an arbitrary elliptic complex. 

 In 1982 Duistermaat and Heckman \cite{PZMR674406} proved the following formula
 \begin{equation}
 \int\limits_M \frac{\omega^n}{n!} e^{-\mu} = \sum\limits_{i} \frac{e^{-\mu(x_i)}}{e(x_i)} ~,
\end{equation}
  where $M$ is a symplectic compact manifold of dimension $2n$ with
  symplectic form $\omega$ and  with a Hamiltonian $U(1)$ action whose
  moment map is $\mu$.  
   Here $x_i$ are the fixed points of the $U(1)$ action and they are assumed
   to be isolated, and $e(x_i)$ is the product of the weights of the $U(1)$ action 
    on the tangent space at $x_i$. Later independently in 1982 Berline and Vergne \cite{PZMR685019} and in 1984 Atiyah and Bott \cite{PZMR721448}
     generalized  the Duistermaat-Heckman formula to the case of a
     general compact manifold $M$ with a $U(1)$ action and 
     an integral $\int \alpha$ of an equivariantly-closed form $\alpha$,
     that is $(d+ \iota_V
     ) \alpha=0$, where $V(x)$ is the vector field corresponding to the  $U(1)$ action. 
       The  Berline-Vergne-Atiyah-Bott formula reads as 
      \begin{equation}
       \int\limits_M \alpha = \sum\limits_{i} \frac{\pi^{n}\alpha_0 (x_i)}{\sqrt{ \det (\partial_\mu V^\nu (x_i))}}~, \label{AB-formula}
      \end{equation}
       where it is assumed that $x_i$ are isolated fixed points of the
       $U(1)$ action, and $\alpha_0$ is the zero-form component of $\alpha$.  
       The Berline-Vergne-Atiyah-Bott formula has multiple generalizations, to the case of non-isolated fixed locus, to supermanifolds,
        to the holomorphic case, etc. The more detailed overview of this formula and its relation to equivariant cohomology is given
         in \volcite{PE}. Here we will concentrate on conceptual issues
         and our discussion is rather schematic.

      Let us review the proof of the Berline-Vergne-Atiyah-Bott formula (\ref{AB-formula}). We will use the language of supergeometry,
      since it is easier to generalize to the infinite dimensional setup. Consider the odd tangent bundle $\Pi TM$ where $x^\mu$ are coordinates 
       on $M$ and $\psi^\mu$ are odd coordinates on the fiber (i.e., they transform as $dx^\mu$). Functions $f(x, \psi)$
        correspond to  differential forms and the integration measure $d^nx~ d^n \psi$ on $\Pi TM$ is canonically defined.
        Assume that there is a $U(1)$ action on compact $M$ with the corresponding vector field $V^\mu(x)\partial_\mu$. Define 
         the following ``supersymmetry  transformations"
      \begin{equation}
        \begin{aligned}
     \delta x^\mu &= \psi^\mu\\
    \delta \psi^\mu &= V^\mu(x) \label{toy-susy}
        \end{aligned}
      \end{equation}
       which correspond to the equivariant differential $d+ \iota_V$. We
       are interested in computation of the integral
          \begin{equation}
    Z  (0)  =    \int\limits_{\Pi TM} \alpha (x, \psi) ~d^nx ~d^n \psi 
      \end{equation}
       for $ \alpha(x,\psi)$ a ``supersymmetric observable", i.e. an
       equivariantly closed form $\delta \alpha (x, \psi)=0$.  We can deform the integral in the following way
         \begin{equation}
          Z(t) = \int\limits_{\Pi TM} \alpha (x, \psi) ~ e^{-t \delta W(x, \psi)}~d^nx ~d^n \psi ~,
          \end{equation}
     where $W(x, \psi)$ is some function. Using the Stokes theorem, one can show that the integral $Z(t)$
       is independent of $t$, provided that $\delta^2 W=0$. For example, we can choose $W= V^\mu g_{\mu\nu} \psi^\nu$ with 
        $g_{\mu\nu}$ being a $U(1)$-invariant metric. If $Z(t)$ is independent of $t$, then we can calculate the original integral 
         at $t=0$ at another value of $t$, in particular we can send $t$ to infinity
            \begin{equation}
        Z(0) = \lim_{t\rightarrow \infty} Z(t) = \lim_{t\rightarrow \infty}  \int\limits_{\Pi TM} \alpha (x, \psi) ~ e^{-t \delta W(x, \psi)}~d^nx ~d^n \psi ~.\label{deriv-AB-formula}
      \end{equation}
      Thus using the saddle point approximation for $Z(t)$ we can calculate the exact value of $Z(0)$. If we choose  
      $W= V^\mu g_{\mu\nu} \psi^\nu$ with the invariant metric and perform the calculation we arrive at the formula (\ref{AB-formula}). 
       Let us outline the main steps of the derivation. In the integral (\ref{deriv-AB-formula})
        \begin{equation}
        \delta W = V^\mu g_{\mu\nu} V^\nu + \partial_\rho (V^\mu g_{\mu\nu}) \psi^\rho \psi^\nu
        \end{equation}
        and thus in the limit $t\rightarrow \infty$ the critical points
        $x_i$ of the $U(1)$ action dominate, $V(x_i)=0$. 
         Let us consider the contribution of one isolated point $x_i$, and for the sake of clarity let's assume that $x_i=0$.
      In the neighbourhood of this critical point $0$, we can rescale coordinates as follows
       \begin{equation}
        \sqrt{t} x = \tilde{x}~,~~~~~ \sqrt{t} \psi = \tilde{\psi}~,
        \end{equation}
      so that the integral expression (\ref{deriv-AB-formula}) becomes
       \begin{equation}
        Z(0)  = \lim_{t\rightarrow \infty}  \int\limits_{\Pi TM} \alpha \left (\frac{\tilde{x}}{\sqrt{t}}, \frac{\tilde{\psi}}{\sqrt{t}} \right 
        ) ~ e^{-t \delta W\left (\frac{\tilde{x}}{\sqrt{t}}, \frac{\tilde{\psi}}{\sqrt{t}} \right )}~d^n
        \tilde{x} ~d^n \tilde{\psi} ~,\label{deriv-2AB}
      \end{equation}
      where we have used the property of the measure on $\Pi TM$,  $d^n x~ d^n \psi = d^n
        \tilde{x} ~d^n \tilde{\psi}$. Now in (\ref{deriv-2AB}) we may 
     keep track of only leading terms which are independent of $t$. In the exponent 
         with $\delta W$ only the quadratic terms are relevant 
               \begin{equation}
          \delta W = H_{\mu\nu} \tilde{x}^\mu \tilde{x}^\nu + S_{\mu\nu} \tilde{\psi}^\mu \tilde{\psi}^\nu~, 
             \end{equation}
              where the concrete form of the matrices $H$ and $S$ is irrelevant. 
   In the limit   $t\rightarrow \infty$ the ``supersymmetry  transformations" (\ref{toy-susy}) are naturally linearized 
   \begin{equation}
        \begin{aligned}
     \delta \tilde{x}^\mu &= \tilde{\psi}^\mu\\
    \delta \tilde{\psi}^\mu &= \partial_\nu V^\mu(0) \tilde{x}^\nu~,\label{linear-susy}
        \end{aligned}
      \end{equation}
      and the condition $\delta^2 W=0$ now implies 
       \begin{equation}
        H_{\mu\nu} = S_{\mu\sigma} \partial_\nu V^\sigma (0)~.\label{rel-S-H}
        \end{equation}
      Now in the integral (\ref{deriv-2AB}) we have to take the limit $t\rightarrow \infty$ and perform the gaussian integral 
       in even and odd coordinates 
       \begin{equation}
       Z(0) =  \alpha (0, 0)  \frac{\pi^{\dim M/2}{\rm Pf }(S)}{\sqrt{\det H}}
      \end{equation}
      and using (\ref{rel-S-H}) we arrive at 
       \begin{equation}
      Z(0) =  \alpha (0, 0)  \frac{\pi^{\dim M/2} }{\sqrt{\det \partial_\mu V^\nu (0)}}~.
      \end{equation}
      If we repeat this calculation for every fixed point, we arrive at the  Berline-Vergne-Atiyah-Bott formula (\ref{AB-formula}). 
       This is the actual proof for a $U(1)$ action on a compact
       $M$. In principle the requirement of a  $U(1)$ action can be
       relaxed to $V$ being Killing vector on a compact $M$, since in
       the derivation we only use the invariance of the metric to construct 
         the appropriate $W$. For non-compact spaces, one can use the  Berline-Vergne-Atiyah-Bott formula as a suitable 
          definition of the integral, for example to introduce the notion of equivariant volume etc. There are many generalizations of 
           the above logic, for example one can construct the holomorphic version of the equivariant differential with the property 
            $\delta^2=0$ etc.

 This setup can be formally generalized to the case where $M$ is  an infinite dimensional manifold. 

Indeed, we can regard this as 
  the definition of the infinite dimensional integral, provided that the formal properties are preserved. 
 However,  in the infinite dimensional case,
  the main challenge is  to make sure that all steps of the formal proof
  can be suitably defined, for example the choice of a suitable 
  $W$ may become a non-trivial problem.  In the infinite dimensional
  situation the matrix  $\partial_\nu V^\mu(0)$ in (\ref{linear-susy}) 
turns into a differential operator
   and the (super)-determinant of this differential operator should be
   defined carefully. 

    The most interesting applications of these ideas come from 
    supersymmetric gauge theories. In this case, one tries to 
     recognise the supersymmetry transformations together with
    the  BRST-symmetry coming from the gauge fixing as some type of 
      equivaraint differential (\ref{toy-susy}) acting on the space of
      fields (an infinite dimensional supermanifold). 

      In the context of the infinite-dimensional path integral, the
      localization construction was first proposed by Witten in his work
      on supersymmetric quantum mechanics \cite{PZMR683171}. In that
      case the infinite dimensional manifold $M$ is the loop space $L X$
      of an ordinary smooth manifold $X$. In the simplest case, the
      $U(1)$ action on $L X$  comes from the rotation of the
      loop. Similar ideas were later applied to two-dimensional
      topological sigma model \cite{PZMR958805} and four dimensional
      topological gauge theory \cite{PZMR953828}.  In the 1990's the ideas
      of localization were widely used in the setup of cohomological
      topological field theories, e.g. see \cite{PZWitten:1992xu} for
      nice applications of these ideas to two-dimensional Yang-Mills theory.
      Further development on supersymmetric localization is related to
      the calculation of Nekrasov's partition function, or equivariant
      Donaldson-Witten theory \cite{PZMR2045303}, based on earlier works
      \cite{PZLosev:1997tp, PZMoore:1997dj, PZLossev:1997bz,
        PZLosev:1995cr}.

 The focus of this volume is on the developments starting from the work \cite{PZPestun:2007rz}, where the exact partition 
   function and the expectation values of Wilson loops for $\mathcal{N}=2$ supersymmetric gauge theories on $S^4$ were calculated.
   In  \cite{PZPestun:2007rz} the 4d $\mathcal{N}=2$ theory was
   placed on $S^4$, preserving 8 supercharges,  and the supersymmetry transformations together with 
    BRST-transformations were recognized as the equivariant differential
    on the space of fields. The zero modes
     were carefully treated by Atiyah-Singer index theorem, and the final result was written as
     a finite-dimensional integral over the Cartan algebra of the Lie
     algebra of the gauge group.  Later this calculation was generalized
     and  extended to other types of supersymmetric theories,  other
     dimensions and geometries.   These exact results provide a unique laboratory 
        for the study of  non-perturbative properties of  gauge theories. 
         Some contributions to this volume provide an overview of the actual localization calculation in concrete dimensions, for 
          concrete class of theories, 
          while other contributions look at the applications of the
          results and discuss their physical and mathematical
          significance.

\section{Localization in diverse dimensions}

In order to apply the localization technique to supersymmetric
theories one needs to resolve a number
 of  technical and conceptual problems. First of all, one needs to define 
 a rigid supersymmetric theory on curved manifolds and understand what geometrical data goes into the 
  construction. The old idea was that rigid supersymmetry on curved manifolds requires an existence 
   of  covariantly constant spinors which would correspond to the parameters in the supersymmetry transformations.
    The next natural generalization would be if the supersymmetry parameters satisfy the Killing spinor equations
    \cite{PZBlau:2000xg}. For example, all spheres admit Killing spinors and thus  supersymmetric gauge theories can 
     be constructed on spheres. However, a more systematic view on supersymmetric rigid theories on 
       curved manifolds has been suggested in \cite{PZFestuccia:2011ws} giving
      background values to auxiliary fields in the supergravity. (More
      recently an approach of topological gravity was explored in \cite{PZImbimbo:2014pla, PZBae:2015eoa}.)
     This approach allows in principle to analyze 
       rigid supersymmetric theories on curved manifolds, although the analysis appears to be increasingly complicated
        as we deal with  higher dimensions and more supersymmetry. At
        the moment we know how 
         to place on a curved manifold the supersymmetric theories,
         which in flat space have four or fewer supercharges,  in
         dimension 2,3 and 4 for both Euclidean and Lorentzian signatures \cite{PZKlare:2012gn,  PZDumitrescu:2012ha, PZCassani:2012ri, PZClosset:2012ru}.  
For other cases only partial results are available. 
    For example, in four dimensions the situation for  theories with eight supercharges remains open, see e.g. \cite{PZKlare:2013dka,
     PZButter:2015tra, PZPestun:2014mja}.  Situation is similar  in five dimensions,  see  e.g. \cite{PZPan:2013uoa, PZImamura:2014ima,
     PZPan:2015nba} and in six dimensions \cite{PZSamtleben:2012ua}; see
   also \cite{PZKuzenko:2015lca, PZKuzenko:2014eqa} in the 
      context  of superspace treatment of rigid supergravity. Thus despite  the
      surge in the activity the full classification of supersymmetric 
       theories on curved manifolds remains an open problem.  Rigid supersymmetric theories on curved manifolds are discussed 
        in  \volcite{DU}.
         
  Moreover, in order to be able to carry the localization calculation explicitly and write the result in closed form,
  we need manifolds with enough symmetries, for example with a rich toric action.
  Again we do not know the full classification of 
    curved manifolds that allow both a toric action and a rigid supersymmetric gauge theory. In 3d  we know how to localize 
    the theories with 4 supercharges on $S^3$, on lens spaces $L_{p}$
    and on  $S^2 \times S^1$.  In 4d the situation becomes more complicated,
     we know how to localize the theories with 8 supercharges on $S^4$ and with 4 supercharges on $S^3 \times S^1$, 
      but the general situation in 4d remains to be understood. In 5d there exists an infinite 
       family of toric Sasaki-Einstein manifold ($S^5$ is one of them)
       for which the result up to non-perturbative contributions can be
       written explicitly for the theories with 8 supercharges.  Notice,
       however, that this is not the most general 5d manifolds which admit the rigid supersymmetry, e.g. a bit separated example is $S^4 \times S^1$.
        In 6d the nearly K\"ahler manifolds (e.g., $S^6$) will allow the theories with 16 supercharges and in 7d the toric Sasaki-Einstein 
         manifolds (e.g., $S^7$) will allow the theories with 16 supercharges. 
 
 The best studied examples are the supersymmetric gauge theories on 
 spheres $S^d$, which we are going to review  briefly since they provide 
  the nice illustration for the general results.
 The first results were obtained 
  for $S^4$ in \cite{PZPestun:2007rz},    for $S^3$ in \cite{PZKapustin:2009kz}, for 
    $S^2$ in   \cite{PZBenini:2012ui, PZDoroud:2012xw}, for  $S^5$  in \cite{PZKallen:2012cs, PZKallen:2012va, PZKim:2012ava} and finally for $S^6$ and $S^7$ were addressed in \cite{PZMinahan:2015jta}. 
      These calculations were generalized and extended to the squashed $S^3$ \cite{PZHama:2011ea, PZImamura:2011wg}, to the squashed $S^4$ \cite{PZHama:2012bg, PZPestun:2014mja},
       the squashed $S^5$ \cite{PZLockhart:2012vp, PZImamura:2012bm, PZKim:2012qf} and the result for the squashed $S^6$ and $S^7$ was already suggested in \cite{PZMinahan:2015jta}. 
 There is also an attempt in \cite{PZMinahan:2015any} to analytically continue
the partition function on $S^d$ to generic complex values of $d$.

 Let us describe the result for different spheres in a uniform fashion. We consider the general case of  squashed spheres. 

The odd and even dimensional spheres $S^{2r-1}$ and $S^{2r}$ lead to  two types of 
special functions called $S_{r}$ and $\Upsilon_r$ that are used to present the result.

 The main building block of these functions  is the 
 multiple inverse Gamma function $\gamma_{r}(x|\ep_1, \dots, \ep_r)$, which is a
 function of a variable $x$ on the complex plane $\mathbb{C}$  and $r$
 complex parameters $\ep_1, \dots, \ep_r$. This function is defined as a $\zeta$-regularized product
\begin{equation}
 \gamma_r (x|\ep_1, ... , \ep_r) = \prod_{n_1, ..., n_r=0}^\infty (x+n_1 \ep_1  +  \cdots + n_r \ep_r),
\end{equation}
The parameters $\ep_i$ should belong to an open half-plane of $\BC$ 
bounded by a real line passing trough the origin.  
  The unrefined  version of  $\gamma_{r}$ is defined as
  \begin{equation}
   \gamma_r (x) = \gamma_r (x|1, ... , 1) = \prod_{k=0}^\infty (x + k)^{\frac{(k+1)(k+2) ... (k+r-1)}{(r-1)!}} ~.
  \end{equation}

The $\Upsilon_r$-function, obtained from the localization on
$S^{2r}$, is defined as
  \begin{equation}
   \Upsilon_r (x|\ep_1, ... , \ep_r) =  \gamma_r \left(x|\ep_1,
     ... , \ep_r \right)  \gamma_r \left(\sum\limits_{i=1}^r \ep_i
   -x|\ep_1, ... , \ep_r \right)^{(-1)^{r}}. \label{def-upsilon}
 \end{equation}
  These functions  form a hierarchy with respect to a shift of $x$ by one
  of $\ep$-parameters 
   \begin{equation}
 \Upsilon_r (x+\ep_i|\ep_1, ... , \ep_i, ... , \ep_r)  = \Upsilon^{-1}_{r-1} (x| \ep_1, ..., \ep_{i-1}, \ep_{i+1}, ..., \ep_r)  
  \Upsilon_r (x |\ep_1, ..., \ep_i , ..., \ep_r)
  \end{equation}
 The  unrefined version of $\Upsilon_r$ is defined 
    as follows
     \begin{equation}
      \Upsilon_r (x) =  \Upsilon_r (x|1, ... , 1) =  \prod_{k \in \mathbb{Z}}  (k+x)^{{\rm sgn}(k+1)\frac{(k+1)(k+2) ... (k+r-1)}{(r-1)!}} ~.
     \end{equation}

 The $S_{r}$-function, called multiple sine, obtained  from localization on
 $S^{2r-1}$, is defined as 
 \begin{equation}
  S_r (x|\ep_1, ... , \ep_r) =  \gamma_r (x|\ep_1, ... ,
  \ep_r)  \gamma_r \left(\sum\limits_{i=1}^r \ep_i
    -x|\ep_1, ... , \ep_r \right)^{(-1)^{r-1}}~.\label{def-multiple-sine}
 \end{equation}
 See \cite{PZNarukawa2004247} for exposition and further
 references.  These functions also  form a hierarchy with respect to a shift of $x$ by one
  o thef $\ep$-parameters 
  \begin{equation}
  S_r (x+ \ep_i |\ep_1, ..., \ep_i , ..., \ep_r) = S^{-1}_{r-1} (x| \ep_1, ..., \ep_{i-1}, \ep_{i+1}, ..., \ep_r)  
  S_r (x |\ep_1, ..., \ep_i , ..., \ep_r).
  \end{equation} 
Notice that  $S_1 (x|\ep) =2 \sin (\frac{\pi x}{\ep})$ is a periodic function. Thus $S_1$ is periodic by itself, $S_2$ is periodic up to $S^{-1}_1$, $S_3$
   is periodic up to $S^{-1}_2$ etc. The unrefined version of multiple sine is defined as
   \begin{equation}
    S_r (x) =  S_r (x| 1, ... ,1) =  \prod_{k \in \mathbb{Z}}  (k+x)^{\frac{(k+1)(k+2) ... (k+r-1)}{(r-1)!}} ~.
    \end{equation}

 The result for a vector multiplet with $4, 8$ and $16$ supercharges
 placed on a sphere $S^2,  S^4$ and $S^6$ respectively is  given in terms of $\Upsilon_r$ functions as follows 
 \begin{equation}
  Z_{S^{2r}} = \int\limits_{{\mathfrak{t}}} d a~  \prod\limits_{w \in R_{\mathrm{ad} \,
    \g}} \Upsilon_{r} (i w \cdot a| \ep ) ~e^{P_{r}(a)} + \cdots~, 
   \label{answer-even}
 \end{equation}
  where the integral is taken over the Cartan subalgebra of the gauge Lie
  algebra $\mathfrak{g}$, the $w$ are weights of the adjoint representation of $\mathfrak{g}$ and 
   $P_r(a)$ is the polynomial in $a$ of degree $r$,
    \begin{equation}
     P_r(a) = \alpha_{r} \Tr (a^{r}) + \cdots + \alpha_2 \Tr (a^2) +
     \alpha_1 \Tr (a). 
    \end{equation} 
 The polynomial $P_{r}(a)$ is coming from the classical action of the
theory. The parameters $\alpha_i$ are related to the Yang-Mills coupling,
the Chern-Simons couplings and the FI couplings. 

The sphere $S^{2r}$ admits $T^r$ action with two fixed points, and the 
   parameters $\ep_1, \dots, \ep_r$ are the squashing parameters for $S^{2r}$ (at the
   same time $\ep_1, \dots, \ep_r $ are equivariant parameters for the
   $T^{r}$  action).

For $S^2$, the dots are non-perturbative contributions coming from 
other localization loci with non-trivial magnetic fluxes (review in \volcite{BL}).
      For $S^4$,  the dots correspond to the contributions of  point-like instantons over 
       the north and south poles computed by the Nekrasov instanton partition
       function (review in \volcite{HO}).  For the case of $S^6$ the
       expression corresponds to maximally supersymmetric theory on
       $S^6$, and   the nature of the dots remains to be understood.  


 The partition function of the vector multiplet with 4, 8, or 16
 supercharges on the odd-dimensional spheres $S^3$, $S^5$ and $S^7$, or $S^{2r-1}$ with $r=2,3,4$, is given by 
  \begin{equation}
  Z_{S^{2r-1}} = \int\limits_{{\bf t}} d a~  \prod\limits_{w \in R_{\mathrm{ad} \,
    \g}} S_{r}
  (i w \cdot a | \ep ) ~e^{P_r(a)} + \cdots~,\label{answer-odd}
 \end{equation}
where now $\ep$-parameters are equivariant parameters of the $T^{r}
\subset SO(2r)$ toric action on $S^{2r-1}$.

    For $S^3$ the dots are absent and the expression (\ref{answer-odd})
    provides the full results for $\mathcal{N}=2$ vector multiplet on
    $S^3$ (review in \volcite{WI}).     For $S^5$ the formula (\ref{answer-odd}) provides the
    result for $\mathcal{N}=1$ vector multiplet (review in \volcite{QZ}).
The theory on $S^7$ is unique and it corresponds to the maximally 
       supersymmetric Yang-Millls in 7d with 16 supercharges. 

     For the case of $S^5$ and $S^7$ the dots are there and they correspond to the contributions around non-trivial connection 
      satisfying certain non-linear PDEs. There are some natural guesses about these corrections, but there are 
      no systematic derivation and no understanding of them, especially for the case of $S^7$.

         Our present discussion can be summarized  in the following table:
  \begin{center}
    \begin{tabular}{ |c|c|c| c|  c | c| c|}
    \hline
 dim   &  multiplet & \#super &special function    & references & derivation \\ \hline
    $~S^2~$  & ~$\mathcal{N}=2$ vector & 4 & $\Upsilon_1(x|\ep)$ ~~
                                       &\cite{PZBenini:2012ui,
                                         PZDoroud:2012xw} & \volcite{BL} \\ \hline
    $S^3$ & $\mathcal{N}=2$ vector & 4&   $S_2(x|\ep)$   &\cite{PZKapustin:2009kz} &\volcite{WI}\\ \hline
    $S^4$ & $\mathcal{N}=2$ vector & 8 & $\Upsilon_2(x|\ep)$   &\cite{PZPestun:2007rz} &\volcite{HO}\\ \hline
     $S^5$ & $\mathcal{N}=1$ vector & 8 &  $S_3(x|\ep)$
                                       &\cite{PZKallen:2012cs,
                                         PZKallen:2012va, PZKim:2012ava}
      &\volcite{QZ} \\ \hline
      $S^6$ & $\mathcal{N}=2$ vector & 16 & $\Upsilon_3(x|\ep)$
                                       &\cite{PZMinahan:2015jta}
                                                    &  \\ \hline
       $S^7$ & $\mathcal{N}=1$ vector & 16& $S_4(x|\ep)$
                                                        &\cite{PZMinahan:2015jta}
                                                                     & \\ \hline
    \end{tabular}
\end{center}

The contribution of matter multiplet (chiral multiplet for theories with $4$
supercharges and hypermultiplets for theories with $8$ supercharges) can
be expressed in terms of the same special functions, see next section. 

   The detailed discussion of the localization calculation on the spheres and other manifolds can be found in different 
   contributions in this volume, 2d is discussed in \volcite{BL}, 3d  in \volcite{WI}, 4d in \volcite{HO},  5d in \volcite{QZ}.

 Next we can schematically explain the above result.  

\subsection{Topological Yang-Mills}

We recall that $\mathcal{N}=1$ super Yang-Mills theory is defined in
dimension $d=3,4,6,10$ and that the algebraic structure of  supersymmetry
transformations is related to an isomorphism that one can establish
between $\BR^{d-2}$ and the famous four division algebras:

  \begin{center}
    \begin{tabular}{ |c|c|c | c| c| c|c| c| c|}
    \hline
SYM   &  algebra & $\mathsf{S}$ & $\dim \mathsf{S}$ & top SYM &
equations & \#equations & quotient\\
\hline
3d & $\BR $  & $S$ &  2&   1d  & --- & 0 &  --- \\
4d & $\BC$ & $S$ &  4 &   2d & $F =0 $ & 1  & Kahler \\
6d & $\mathbb{H} $ &  $S^{+}\otimes \BC^2 $&  8 & 4d & $F = - \star F$
&3 & hyperKahler \\
10d & $\BO $ &  $S^{+}$ &  16 & 8d & $F = - \star (F \wedge \Omega)$ & 7
& octonionic 
 \\
\hline
    \end{tabular}
\end{center}

In this table $S$ denotes the $2^{\lfloor d/2 \rfloor}$-dimensional Dirac spinor representation of
 $Spin(d)$ group. The $S^{+}$ denotes the chiral (Weyl) spinor
 representation of $Spin(d)$. In all cases, one uses Majorana spinors in
 Lorenzian signature, or \emph{holomorphic Dirac}\footnote{This means
   that the spinors $\psi$ in Euclidean signature are taken to be complex, but
   algebraically speaking, only $\psi$ appears in the theory but not its complex
   conjugate $\bar \psi$} spinors in Euclidean signature. Notice the
 peculiarity of the 6d case where one uses chiral $Sp(1)$-doublet spinors with
 $\BC^2$ being the fundamental representation of the $Sp(1) \simeq
 \SU(2)$ R-symmetry, and that in the 10d case one uses a single copy
of the chiral spinor representation.  The number $\dim \mathsf{S}$ is
often referred as the \emph{the number of the supercharges in the
  theory}.

Also, it is well known that the $\mathcal{N}=1$  
under the dimensional reduction to the dimension $d-2$ produces the
`topological' SYM which localizes to the solutions of certain
first order (BPS type) elliptic equations on the gauge field strength of
the curvature listed in the table. 
The 1d topYM is, of
course, the trivial theory, with empty equations, since there is no room
for the curvature 2-form in a 1-dimensional theory.
The equation for 2d topYM is the equation of zero curvature, for 4d
topYM it is the instanton equation of self-dual curvature (defined by
the conformal structure on the 4d manifold), 
and for 8d topYM it is the equation of the octonionic instanton (defined by the Hodge star $\star$ operator and the
Cayley 4-form $\Omega$ on a  $Spin(7)$-holonomy 8d manifold).

The corresponding linearized complexes are

  \begin{equation}
    \begin{tabular}{ c | c | c }
topYM & linearized complex & fiber dimensions\\
\hline
$\BR$: 1d      & $ \Omega^{0} \to \Omega^{1} $ & $\mathbf{1} \to \mathbf{1}$ \\
$\BC$: 2d      & $ \Omega^{0} \to \Omega^{1}  \to \Omega^{2}$& $\mathbf{1} \to
 \mathbf{2 } \to \mathbf{1}$ \\
$\mathbb{H}$: 4d      & $ \Omega^{0} \to \Omega^{1} \to \Omega^{2}_{+}$ & $\mathbf{1}
\to \mathbf{4} \to \mathbf{3}$ \\
$\BO$: 8d      & $ \Omega^{0} \to \Omega^{1} \to \Omega^{2}_{\mathrm{oct}}$ & $\mathbf{1} \to \mathbf{8} \to \mathbf{7} $\\
    \end{tabular}
\label{eq:complexes}
\end{equation}

Here $\Omega^{p}$ is a shorthand for $\Omega^{p}(X) \otimes \mathrm{ad}
\, \mathfrak{g}$, that is the space of $\g$-valued differential
$p$-forms on $X$,  where $\mathfrak{g}$ is the Lie algebra of the gauge
group and $X$ is the space-time manifold. 
In the 4d theory the space $\Omega_{+}^2$ denotes the space of
self-dual 2-forms that satisfy the instanton equation $F = - \star F$,
and in the 8d
theory the space $\Omega^{2}_{\mathrm{oct}}$ is the space of  2-forms
that satisfy the octonionic instanton equation $F = - \star (F \wedge \Omega)$.

In these complexes, the first term $\Omega^{0}$ describes the tangent 
space to the infinite-dimensional  group of gauge transformations on $X$, the second term
$\Omega^{1}$ describes the tangent space to the affine space of gauge
connections on $X$, and the last term ($\Omega^2$ for 2d, $\Omega^2_{+}$
for 4d, $\Omega^{2}_{\mathrm{oct}}$ for 8d) describes that space where the
equations are valued.

If space-time $X$ is invariant under an isometry group $T$, 
the topological YM can be treated equivariantly with respect to the
$T$-action. The prototypical case is
 the equivariant Donaldson-Witten theory, or 4d topYM in
 $\Omega$-background
defined on $\BR^{4}$ equivariantly with respect to $T = SO(4)$, generating
the Nekrasov partition  function~\cite{PZMR2045303}.  Special functions,
like the $\Upsilon$-function defined by
 infinite products like (\ref{def-upsilon}) are
infinite-dimensional versions of the equivariant Euler class of the
tangent bundle to the space of all fields appearing
after localization of the path integral by Atiyah-Bott fixed point
formula (see \volcite{PE} section 8.1 for more details). The equivariant Euler class can be
determined by computing first the equivariant Chern class (index) of the
linearized complex describing the tangent space of the topological YM
theory. The $T$-equivariant Chern class (index) for the equation elliptic
complex 
\begin{equation}
D: \dots \to \Gamma(E_k, X) \to \Gamma(E_{k +1}, X) \to \dots   
\end{equation}
 on space $X$ made from sections of vector bundles $E_\bullet$, can be conveniently computed by the Atiyah-Singer index theorem
\begin{equation}
 \ind_{T}(D) = \sum_{x \in X^{T}} \frac{ \sum_{k} (-1)^k
    \ch_T(E_k)|_x }{ \det_{T_{x}X }(1 - t^{-1})}
\end{equation}
where $X^{T}$ is the fixed point set of $T$ on $X$ (see \volcite{PE} section 11.1 for details).

\subsection{Even dimensions}

First we will apply the Atiyah-Singer index theorem (review in
\volcite{PE} section 11.1)  for the \emph{complexified} complexes (\ref{eq:complexes})
on $X = \BR^{d}$ for $d=2,4,8$ topological YM with respect to the natural $SO(d)$
equivariant action on $\BR^{d}$ with fixed point $x = 0$. 

For $d=2r$ and $r=1,2,4$ we pick the Cartan torus $T^{r} = U(1)^{r}$ in the $SO(2r)$
with parameters $(t_1, \dots, t_r) \in U(1)^{r}$. The denominator in the
Atiyah-Singer index theorem is 
\begin{equation}
  \det(1-t^{-1})|_{\BR^{2r}} = \prod_{s=1}^{n} (1 - t_s)(1-t_s^{-1})
\end{equation}

The numerator is obtained by computing the graded trace over the fiber
of the equation complex at the fixed point $x=0$.  


For equivariant 2d topYM on $\BR^{2}$ (coming from SYM with 4 supercharges): 
  \begin{equation}
 \ind_{T}(D, \BR^{2}, \Omega_{\BC}^{0} \to \Omega_{\BC}^{1}  \to \Omega_{\BC}^{2})_{\mathrm{2d}} =
\frac{ 1 - (t_1 + t_1^{-1}) + 1}{ (1 - t_1)(1-t_1^{-1})} = \frac{1}{1 -t_1} + \frac{1}{1 -t_1^{-1}}
\label{eq:2dindex}
 \end{equation}

For equivariant 4d topYM on $\BR^{4}$ (coming from SYM with 8 supercharges):
  \begin{multline}
 \ind_{T}(D, \BR^{4}, \Omega_{\BC}^{0} \to \Omega_{\BC}^{1}  \to \Omega_{\BC}^{2+})_{\mathrm{4d}} =
 \frac{ 1 - (t_1 + t_1^{-1} + t_2 + t_2^{-1}) + (1 + t_1 t_2 +
  t_1^{-1} t_2^{-1})}{ (1-t_1)(1-t_1^{-1})(1-t_2)(1-t_2^{-1})}\\
= \frac{1}{(1-t_1)(1-t_2)} + \frac{1}{(1-t_1^{-1})(1-t_2^{-1})}
\label{eq:4dindex}
 \end{multline}

For equivariant 8d topYM on $\BR^8$ (coming from SYM with 16
supercharges), to preserve the Cayley form and
  the octonionic equations coming from the $Spin(7)$ structure, the 4
  parameters $(t_1, t_2, t_3, t_4)$  should satisfy the constraint $t_1
  t_2 t_3 t_4  = 1$. The weights on 7-dimensional bundle, whose sections 
are  $\Omega_{\mathrm{oct},\BC}^{2}$, can be
computed from the weights of the chiral spinor bundle $S^{+}$ modulo the
trivial bundle. The
chiral spinor bundle $S^{+}$ can be identified (after a choice of
complex structure on $X$) as $S^{+} \simeq
(\oplus_{p=0}^{2} \Lambda^{2p} T_{X}^{0,1})\otimes K^{\frac 1 2}$ where $K$
is the canonical bundle on $X = \BR^{8} \simeq \BC^{4}$ equivariantly
trivial with respect to the $T^3$ action parametrized by $(t_1, t_2,
t_3, t_4)$ with $t_1 t_2 t_3 t_4 = 1$. 
Then
  \begin{multline}
\ind_{T}(D, \BR^{8}, \Omega_{\BC}^{0} \to \Omega_{\BC}^{1}  \to
\Omega_{\mathrm{oct},\BC}^{2})_{\mathrm{8d}} = \frac{ (1 -(\sum_{s=1}^{4} (t_s + t_s^{-1}))
 + (1 + \sum_{ 1 \leq r < s \leq
    4} t_r t_s )}
{ \prod_{s=1}^{4} (1 - t_s)(1-t_s^{-1}) } =\\
=  \frac{1}{(1 -t_1)(1-t_2)(1-t_3)(1- t_4)}, \qquad t_1 t_2 t_3 t_4 = 1
\label{eq:8dindex}
 \end{multline}

It is also interesting to consider the dimensional reduction of the 8d
topYM (coming from the SYM with 16 supercharges) to the 6d theory. The
numerator in the index is computed in the same way as
(\ref{eq:8dindex}), but the denominator is changed to the 6d
determinant, hence we find 
\begin{multline}
 \ind_{T}(D, \BR^{6}, \Omega_{\BC}^{0} \to \Omega_{\BC}^{1}  \to
\Omega_{\BC}^{2,\mathrm{oct}})_{\mathrm{6d\,reduction}} = \frac{ (1 -(\sum_{s=1}^{4} (t_s + t_s^{-1}))
 + (1 + \sum_{ 1 \leq r < s \leq
    4} t_r t_s )}
{ \prod_{s=1}^{3} (1 - t_s)(1-t_s^{-1}) } =\\
=  \frac{1-t_4^{-1}}{(1 -t_1)(1-t_2)(1-t_3)}  \stackrel{ t_1 t_2 t_3 t_4
  = 1} = \\
= \frac{1}{(1-t_1)(1-t_2)(1-t_3)}+\frac{1}{(1-t_1^{-1})(1-t_2^{-1})(1-t_3^{-1})}
\label{eq:6dindex}
\end{multline}

From equations (\ref{eq:2dindex})(\ref{eq:4dindex})(\ref{eq:6dindex}) we
see that the index for the complexified vector multiplet of the 2d theory (4 supercharges), 4d theory (8 supercharges) and 6d theory
(16 supercharges)  on $\BR^{2r}$ can be uniformly written in the form 
\begin{multline}
  \ind_{T}(D, \BR^{2r},\mathrm{vector}_{\BC}) =  \frac{ 1 + (-1)^{r} \prod_{s=1}^{r} t_s}{
    \prod_{s=1}^{r}(1-t_s)} = \frac{ 1}{
    \prod_{s=1}^{r}(1-t_s)} + \frac{ 1}{
    \prod_{s=1}^{r}(1-t_s^{-1})}  
\qquad  r= 1,2,3
\label{eq:index123}
\end{multline} 

Hence,  the equivariant index of the complexified vector multiplet in 2,4 and 6
dimensions on flat space is equivalent to the index of the Dolbeault complex plus its
dual, because (see review in \volcite{PE} section 9)
\begin{equation}
  \ind_{T}(\bar \partial, \BC^{2r}, \Omega^{0, \bullet}) = \frac 1
  {\prod_{s=1}^{r} (1 - t_s^{-1})}
\end{equation}

The vector multiplet is in a real representation of the equivariant group:
 each non-zero weight eigenspace appears together with its
dual. Generally, the index of a real representation has the form
\begin{equation}
  f(t_1, \dots, t_r) + f(t^{-1}_1, \dots, t_{r}^{-1})
\end{equation}

The equivariant Euler class in the denominator of the Atiyah-Bott
localization formula (\volcite{PE} section 8.1 and section 12) is
defined as the Pfaffian rather then the determinant, hence each pair of
terms in the equivariant index, describing
a weight space and its dual, corresponds to a single weight
factor in the equivariant Euler class. The choice between two opposite 
weights leads to a sign issue, which depends on
the choice of the orientation on the infinite-dimensional space of all
field modes.   A careful treatment leads to interesting sign factors
discussed in details for example in \volcite{BL}. 

A natural choice of orientation leads to the holomorphic projection of
the vector multiplet index \eqref{eq:index123}  in 2, 4 and 6 dimensions
by picking only the first term in \eqref{eq:index123} so that
\begin{equation}
  \ind_{T}(D, \BR^{2r},\mathrm{vector}_{\BC})_{\mathrm{hol}} =  \frac{ 1}{
    \prod_{s=1}^{r}(1-t_s)}  
\qquad  r= 1,2,3
\label{eq:index123hol}
\end{equation}

The supersymmetric Yang-Mills with 4, 8 and 16 supercharges can be put
on the spheres $S^2$,  $S^4$ and $S^6$ as was done in
\cite{PZPestun:2007rz}, \cite{PZBenini:2012ui}, \cite{PZDoroud:2012xw},
\cite{PZMinahan:2015jta} and reviewed in \volcite{BL} and \volcite{HO}.

A certain generator $Q_{\ep}$ of the global superconformal group can be used
for the localization computation. This generator $Q_{\ep}$ is
represented by a conformal Killing spinor $\ep$ on a sphere $S^{2r}$, and satisfies $Q_{\ep}^2
= R$ where $R$ is a rotation isometry.  There are two fixed points of
$R$ on an even-dimensional sphere, usually called the north and the south poles. It turns out that the
equivariant elliptic complex of equations,  describing the equations of the topological
YM, is replaced by a certain equivariant transversally elliptic complex
of equations. Near the north pole this complex is approximated by the
equivariant topological YM theory (theory in $\Omega$-background), and
near the south pole by its conjugate. 

The index of the transversally elliptic operator can be computed by the
Atiyah-Singer theorem, see for the complete treatement
\cite{PZAtiyah1974}, application \cite{PZPestun:2007rz},  \volcite{PE} or
\volcite{HO}. The result is that the index is contributed by the two
fixed point on the sphere $S^{2r}$, with a particular choice of the
distribution associated to the rational function, in other words with a particular
choice of expansion in positive or negative powers of $t_s$, denoted
by $[]_{+}$ or $[]_{-}$ respectively (see \volcite{PE} section 11.1):
\begin{equation}
  \ind_{T}(D, S^{2r},\mathrm{vector}_{\BC})_{\mathrm{hol}} =  \left[\frac{ 1}{
    \prod_{s=1}^{r}(1-t_s)}  \right]_{+} + \left[\frac{ 1}{
    \prod_{s=1}^{r}(1-t_s)}  \right]_{-}
\qquad  r= 1,2,3
\label{eq:index123holsphere}
\end{equation}   

So far we have computed only the space-time geometrical part of the
index. Now, suppose that the multiplet is tensored with a representation
of a group $G$ (like the gauge symmetry, R-symmetry or flavour
symmetry), and let $L_{\xi} \simeq \BC$ be a complex eigenspace in
representation of $G$  with eigenweight $\xi = e^{i x}$. Then 

\begin{equation}
  \ind_{T \times G}(D, S^{2r},\mathrm{vector}_{\BC} \otimes
  L_\xi)_{\mathrm{hol}} =    \left[\frac{ \xi}{
    \prod_{s=1}^{r}(1-t_s)}  \right]_{+} +  \left[\frac{ \xi}{
    \prod_{s=1}^{r}(1-t_s)}  \right]_{-}
\label{eq:Upsilon-character}
\end{equation}

Now let $\ep_s$ and $x$ be the Lie algebra parameters associated with
the group parameters $t_s$ and $\xi$ as
\begin{equation}
  t_s = \exp( i \ep_s), \qquad  \xi = \exp( i x)
\end{equation}

By definition, let  $\Upsilon_{r}(x | \ep)$  be the equivariant Euler class
(Pfaffian) of the graded vector space of fields of a vector multiplet on
$S^{2r}$ with the character
(index) defined by \eqref{eq:Upsilon-character} 
\begin{equation}
 \Upsilon_r(x|\ep) = \mathrm{eu}_{T \times G}(D,
  S^{2r},\mathrm{vector}_{\BC} \otimes L_\xi )_{\mathrm{hol}}|_{t_s =
    e^{i \ep_s}, \xi = e^{i x} } 
\label{eq:Upsilon-vector}
\end{equation}

Explicitly, converting the infinite Taylor sum series of
\eqref{eq:Upsilon-character} 
\begin{equation}
  \left[\frac{ \xi}{
    \prod_{s=1}^{r}(1-t_s)}  \right]_{+} +  \left[\frac{ \xi}{
    \prod_{s=1}^{r}(1-t_s)}  \right]_{-}=\sum_{n_1=0, \cdots, n_r =
  0}^{\infty} \xi (t_1^{n_1} \dots t_r^{n_r} +  (-1)^{r} t_1^{-1-n_1} \cdots t_r^{-1-n_r})
\end{equation}
into the product of weights we find the infinite-product definition  of the
$\Upsilon_{r}(x|\ep)$ function
\begin{equation}
  \Upsilon_r(x|s) \stackrel{reg}{=} \prod_{n_1=0, \dots, n_r=0}^{\infty} \left(x +
  \sum_{s=1}^{r} n_s \ep_s\right)\left(\ep  - x +   \sum_{s=1}^{r} n_s \ep_s\right)^{(-1)^{r}}
\label{eq:Upsilon-product}
\end{equation}
where $\stackrel{reg}{=}$ denotes Weierstrass or $\zeta$-function
regularization and 
\begin{equation}
  \ep = \ep_1 + \dots +\ep_{r}
\end{equation}

The analysis for the scalar multiplet (the chiral multiplet in 2d for
the theory with 4 supercharges or the hypermultiplet in 4d for the
theory with 8 supercharges) is similar. On equivariant $\BR^{2r}$ the
corresponding complex for the scalar multiplet is the Dirac operator $S^{+} \to S^{-}$,
which differs from the Dolbeault complex by the twist by the square root
of the canonical bundle, hence 
\begin{equation}
    \ind_{T}(D, \BR^{2r},\mathrm{scalar})_{\mathrm{hol}} =  -\frac{
      \prod_{s=1}^{r} t_s^{\frac 1 2} }{
    \prod_{s=1}^{r}(1-t_s)}  
\qquad  r= 1,2
\label{eq:indexscalarR}
\end{equation}

On the sphere $S^{2r}$, again, one takes the contribution from the north and the
south pole approximated locally by $\BR^{2r}$ with opposite orientations, and gets
\begin{equation}
    \ind_{T}(D, S^{2r},\mathrm{scalar})_{\mathrm{hol}} =  -\left[\frac{
      \prod_{s=1}^{r} t_s^{\frac 1 2} }{
    \prod_{s=1}^{r}(1-t_s)}  \right]_{+} - \left[\frac{
      \prod_{s=1}^{r} t_s^{\frac 1 2} }{
    \prod_{s=1}^{r}(1-t_s)}  \right]_{-} \qquad  r= 1,2
\label{eq:indexscalarS}
\end{equation}

Hence, the equivariant Euler class of the graded space of sections of the
scalar multiplet is obtained simply by a shift of the argument of the
$\Upsilon$-function and inversion
\begin{equation}
      \mathrm{eu}_{T \times G}(D, S^{2r},\mathrm{scalar}\otimes L_{\xi})_{\mathrm{hol}}|_{t_s =
    e^{i \ep_s}, w = e^{i x} } = \Upsilon_{r} \left(x + \frac \ep 2\right)^{-1}
\label{eq:Upsilon-scalar}
\end{equation}

As computed in \cite{PZPestun:2007rz}, \cite{PZBenini:2012ui}, \cite{PZDoroud:2012xw},
\cite{PZMinahan:2015jta} and reviewed in \volcite{BL} and \volcite{HO},
the localization by the Atiyah-Bott formula brings
the partition function of  supersymmetric Yang-Mills with 4, 8 and 16 supercharges 
on the spheres $S^2$,  $S^4$ and  $S^6$ to the form of an integral
over the imaginary line contour in the complexified Lie algebra of the
Cartan torus of the gauge group (the zero mode of one of the scalar fields in the vector
multiplet). The integrand is a product of the classical factor induced from
the classical action and the determinant factor (the inverse of the equivariant Euler
class of the tangent space to the space of fields) which has been
computed above in terms of the $\Upsilon_r$-function. Hence, for $r=1,2,3$
we get perturbatively exact result of the partition function in the form
of a finite-dimensional integral over the Cartan subalgebra of the Lie
algebra of the gauge group (generalized matrix model) 
\begin{equation}
  Z_{S^{2r}, \mathrm{pert}} = \int_{\mathfrak{t}_{G}} da \frac{\prod_{w \in R_{\mathrm{ad} \,
    \g} } \Upsilon_{r}( i w \cdot a|\ep)}{ \prod_{w \in R_{G
        \times F}} \Upsilon_{r}( i w \cdot
  (a,m)+ \frac{\ep}{2}|\ep)} \, e^{P(a)} 
\label{eq:ZpertSeven}
\end{equation}

Hence  $Z_{S^{2r}, \mathrm{pert}}$ is the contribution to the partition function of the
trivial localization locus (all fields vanish except the zero mode $a$ of one of the scalars
of the vector multiplet and some auxliary fields). The $Z_{S^{2r},\mathrm{pert}}$  does not include
the non-perturbative contributions. The factor $e^{P(a)}$
is induced by the classical action evaluated at the localization
locus. The product of $\Upsilon_r$-functions in the numerator comes from the
vector multiplet and it runs over the weights of the adjoint
representation. The product of $\Upsilon_r$-functions in the denominator
comes from the scalar multiplet (chiral or hyper), and it runs over the
weights of a complex
representation $R_{G}$ of the gauge group $G$ in which the scalar multiplet
transforms. In addition, by taking
the matter fields multiplets to be in a representation of a flavor symmetry
$F$, the mass parameters $m \in \mathfrak{t}_{F}$ can be introduced
naturally. For $r=3$ the denominator is empty, because the 6d gauge theory with 16
supercharges is formed only from the gauge vector multiplet.

The non-perturbative contributions come from other localization
loci,  such as magnetic fluxes on  $S^2$,  or instantons  on  $S^4$,
and their effect modifies the equivariant Euler classes presented as
$\Upsilon_{r}$-factors in \eqref{eq:ZpertSeven} by certain rational
factors. The 4d non-perturbative contributions are captured by
fusion of Nekrasov instanton partition function with its conjugate
\cite{PZMR2045303, PZPestun:2007rz}.  See 2d details in \volcite{BL} and 4d details in
\volcite{HO}.

Much before localization results on gauge theory on $S^4$ were obtained,
the $\Upsilon_2$ function prominently appeared in
Zamolodchikov-Zamolodchikov paper   \cite{PZZamolodchikov:1995aa} on structure functions
of 2d Liouville CFT. The coincidence was one of the key observations by
Alday-Gaiotto-Tachikawa \cite{PZAlday:2009aq} that led
 to a remarkable 2d/4d correspondence (AGT) between
correlators in Liouville (Toda) theory and gauge theory partition functions on
$S^4$, see review in \volcite{TA}.

\subsection{Odd dimensions}

Next we discuss the odd dimensional spheres (in principle, this discussion is 
  applicable for any simply connected Sasaki-Einstein manifold, i.e. the
  manifold $X$ admits at least two Killing spinors).  After  field
  redefinitions, which involve the Killing spinors,  
 the integration space for odd dimensional  supersymmetric gauge theories with the gauged 
  fixing fields can be represented as 
  the following spaces
  \begin{equation}
  \begin{aligned}
&  {\rm 3d:} &\qquad & {\mathcal A} (X, \g) \times \Pi \Omega^0 (X, \g) \times \Pi \Omega^0 (X, \g) \times \Pi \Omega^0 (X, \g)\\
&  {\rm 5d:} &\qquad &  {\mathcal A} (X, \g) \times \Pi \Omega_H^{2,+} (X, \g) \times \Pi \Omega^0 (X, \g) \times \Pi \Omega^0 (X, \g)\\
&  {\rm 7d:} &\qquad &  {\mathcal A} (X, \g) \times \Omega^{3,0}_H (X, \g)
  \times \Pi \Omega_H^{2,+} (X, \g) \times \Pi \Omega^0 (X, \g) \times \Pi \Omega^0 (X, \g)
     \end{aligned}
   \end{equation}
 where in all cases there are common
  last two factors $\Pi \Omega^0 (X, \g) \times \Pi \Omega^0 (X, \g)$ coming from the gauge fixing.  
   The space ${\mathcal A} (X, \g)$ is the space of connections on $X$ with the Lie algebra $\g$. The Sasaki-Einstein manifold
    is a  contact manifold and the differential forms can be naturally decomposed into vertical and horizontal forms using the Reeb vector field $R$
     and the contact form $\kappa$. The horizontal plane admits  a complex structure  and thus the horizontal forms can be decomposed 
      further into $(p,q)$-forms.  For two forms we define the space
      $\Omega_H^{2,+}$ as $(2,0)$-forms plus $(0,2)$-forms plus forms proportional 
       to $d\kappa$. Thus for 5d $\Omega_H^{2,+}$ is the space of
       standard self-dual forms in four dimensions (rank 3 bundle), 
        and for 7d forms in $\Omega_H^{2,+}$ obey the
 hermitian Yang-Mills conditions in six dimensions (rank 7 bundle: 3
 complex components and 1 real). 
      By just counting  degrees of freedom
      one can check that the 3d case corresponds to an $\mathcal{N}=2$ vector 
      multiplet (4 supercharges), the 5d case to an $\mathcal{N}=1$ vector multiplet (8 supercharges) and 7d to $\mathcal{N}=1$ maximally supersymmetric 
       theory (16 supercharges).  The supersymmetry square $Q_{\ep}^2$, which acts on this
       space, is given by the sum of Lie derivative along the Reeb vector field $R$ and 
       constant gauge transformations: $Q_{\ep}^2 = {\mathcal L}_R +
       ad_a$. Around the trivial connection,  after some cancelations,  the problem boils down to 
         the calculation of the following superdeterminant 
     \begin{equation}
   Z_{S^{2r-1}} = \int\limits_{{\mathfrak{t}_{G}}} d a~ {\rm sdet}_{\Omega_H^{(\bullet, 0)}(X, \g)} ({\mathcal L}_R + ad_a)  ~e^{P_r(a)} + \cdots~,
\label{eq:Zsphere-odd}
     \end{equation}
  and this is a uniform description for  Sasaki-Einstein manifolds in 3d, 5d and 7d\@.  
   In 3d the only simply connected Sasaki-Einstein 
   manifold is $S^3$, while in 5d and 7d there are many examples of  simply connected Sasaki-Einstein 
   manifolds (there is a rich class of the toric  Sasaki-Einstein  manifolds). The determinant can be calculated in many 
    alternative ways, and the result depends on $X$. 

If $X$ is a sphere $S^{2r-1}$,  the determinant in
\eqref{eq:Zsphere-odd}, equivalently, the inverse equivariant Euler
class of the normal bundle to the localization locus in the space of all
fields, can be computed from the equivariant Chern character, or the
index, of a certain transversally elliptic operator $D =  \pi^{*}
\bar \partial$ induced from the Dobeault operator $\bar \partial$ 
by the Hopf fibration projection $\pi: S^{2r-1} \to \CP^{r-1}$. 

The index, or equivariant Chern character, is easy to compute by the
Aityah-Singer fixed point theorem (see the details in \volcite{PE} section 11.2). The result is 
\begin{equation}
  \ind_{T  }(D, S^{2r-1}) = \sum_{n=-\infty}^{\infty}
  \ind_{T}(\bar \partial , \CP^{r-1}, \mathcal{O}(n))  =
\left[\frac{1}{\prod_{k=1}^{r}(1 - t_k)}\right]_{+} + \left[\frac{ (-1)^{r-1} t_1^{-1} \dots t_r^{-1}}{\prod_{k=1}^{r}(1 - t_k^{-1})}\right]_{-}
\end{equation}

Converting the additive equivariant Chern character to the multiplicative
equivariant Euler character, we find the definition of the multiple sine function
\begin{equation}
 S_r(x|\ep) = \mathrm{eu}_{T \times G}(  S^{2r-1}, D \otimes L_{\xi})_{\mathrm{hol}}|_{t_s =
    e^{i \ep_s}, \xi = e^{i x} } 
\label{eq:Sine-vector}
\end{equation}
where $L_{\xi}$ is a 1-dimensional complex eigenspace with character $\xi$. 
Explicitly
\begin{equation}
  S_r(x|\ep)\stackrel{reg}{=} \prod_{n_1=0, \dots, n_r=0}^{\infty} \left(x +
  \sum_{s=1}^{r} n_s \ep_s\right)\left(\ep  - x +   \sum_{s=1}^{r} n_s \ep_s\right)^{(-1)^{r-1}}
\label{eq:S-product}
\end{equation}
and this leads to the formula \eqref{answer-odd} for the perturbative
part of the partition function of  a vector multiplet on $S^{2r-1}$. 

For $r=2,3$ we can also treat a scalar supermultiplet (a chiral multiplet
for the theory with 4 supercharges or a hypermultiplet for the theory with
8 supercharges). The corresponding complex is described by an elliptic
operator $\pi^{*} \slashed{D}$ for $\pi: S^{2r-1} \to \CP^{r-1}$, where
$\slashed{D}$ is the Dirac operator $S^{+} \to S^{-}$  on $\CP^{r-1}$. The
Dirac complex is isomorphic to the Dolbeault complex by a twist by a
square root of the canonical bundle. Because of the opposite
statistics, there is also an overall sign factor like in
\eqref{eq:indexscalarS}.  

Finally, the contribution of both vector
multiplet in representation $R_{\mathrm{ad}\, \g}$ and scalar multiplet
in representation $R_{G \times F}$ to the perturbative part of the
partition function is computed by the finite-dimensional integral over the localizationl
locus $\mathfrak{t}_{G}$ with the following integrand made of $S_{r}$ functions
\begin{equation}
  Z_{S^{2r-1}, \mathrm{pert}} = \int_{\mathfrak{t}_{G}} da \frac{\prod_{w \in R_{\mathrm{ad} \,
    \g} } S_{r}( i w \cdot a|\ep)}{ \prod_{w \in R_{G
        \times F}} S_{r}( i w \cdot
  (a,m)+ \frac{\ep}{2}|\ep)} \, e^{P(a)} 
\label{eq:ZpertSodd}
\end{equation}
Here $F$ is a possible flavor group of symmetry, and $m \in
\mathfrak{t}_{F}$ is a mass parameter.

For  reviews of 3d localization see \volcite{WI}, \volcite{MA}, \volcite{PU},
\volcite{DI}  and for reviews of  5d localization see \volcite{QZ}, \volcite{MI},
\volcite{PA}.

 The case of $S^n \times S^1$ is  built from the trigonometric version of $S^n$-result.  

The trigonometric  version of the $\Upsilon_r$-function \eqref{eq:Upsilon-product} is given by 
    \begin{equation}
  H_r (x|\ep_1, ... , \ep_r) = \prod\limits_{n_1, ... , n_r=0}^\infty \Big(1 - e^{2\pi i x} e^{2\pi i \vec{n} \vec{\ep}}  \Big)
    \Big(1 - e^{2\pi i (\sum\limits_{i=1}^r \ep_i -  x)} e^{2\pi i \vec{n} \vec{\ep}}  \Big)^{(-1)^{r}}~.
   \end{equation}

The trigonometric version of 
the  multiple sine function $S_{r}$  \eqref{eq:S-product} is given by the multiple elliptic gamma function
  \begin{equation}
   G_r (x|\ep_1, ... , \ep_r) = \prod\limits_{n_1, ... , n_r=0}^\infty \Big(1 - e^{2\pi i x} e^{2\pi i \vec{n} \vec{\ep}}  \Big)
    \Big(1 - e^{2\pi i (\sum\limits_{i=1}^r \ep_i - x)} e^{2\pi i \vec{n} \vec{\ep}}  \Big)^{(-1)^{r-1}}~.
   \end{equation}
  where $G_1$ corresponds to the $\theta$-function, $G_2$ corresponds to
  the elliptic gamma function.


  The  partition function on $S^r \times S^1$ has an interpretation as a
  supersymmetric index, namely a graded trace over the Hilbert space. 
The review of supersymmetric index in 2d is in \volcite{BL}, in 4d is in
\volcite{RR} and in 6d is in \volcite{KL}.
 

\section{Applications of the localization technique}

The localization technique can be applied only to a very restricted set
of supersymmetric observables, e.g. partition functions,
supersymmetric Wilson loops etc. Unfortunately,  the localization
technique does not allow us to calculate  correlators of  generic
local operators. However,  the supersymmetric localization offers a unique opportunity
to study the full non-perturbative answer for these restricted class of
observables and this is a powerful tool to inspect interacting quantum
field theory.  As one can see from the previous section, the
localization results are given in terms of complicated finite
dimensional integrals. Thus one has to develop  techniques to study
these integrals and learn how to deduce the relevant physical and
mathematical information.  Some of the reviews in this volume are
dedicated to the study of the localization results (sometimes in
various limits) and to the applications of these results in physics
and mathematics.

The original motivation of \cite{PZPestun:2007rz} was to prove  the
Erickson-Semenoff-Zarembo and Drukker-Gross conjecture, 
 which expresses the expectation value of supersymmetric circular Wilson loop operators in  
 $\mathcal{N} = 4$ supersymmetric Yang-Mills theory in terms of a
 Gaussian matrix model, see review in \volcite{ZA}. 
  This conjecture was actively used for checks of AdS/CFT
  correspondence. After more general localization results became available,
  they were also used for  stronger tests of  AdS/CFT.

  On the AdS side, it is relatively easy to perform the calculation, since
  it requires only classical supergravity. However, on the gauge theory
  side, we need the full non-perturbative result in order to be able to
  compare it with the supergravity calculation.  The localization
  technique offers us a unique opportunity for non-perturbative
  checks of AdS/CFT correspondence.  A number of reviews are devoted to
  the use of localization for AdS/CFT correspondence: for
  AdS$_4$/CFT$_3$ see review in  \volcite{MA} and \volcite{PU}, for AdS$_5$/CFT$_4$ see
  review in \volcite{ZA}, for AdS$_{7}$/CFT$_6$ see review in \volcite{MI} and \volcite
  {KL}.  The localization results for spheres (\ref{eq:ZpertSeven}) and
  (\ref{eq:ZpertSodd}) gave rise to new matrix models which had not been
  investigated  before.  One of the main problems is to find out how the free
  energy (the logarithm of the partition function) scales in the large
  $N$-limit. In 3d there is an interesting scaling $N^{3/2}$, and
  the analysis of the partition function on $S^3$ for the ABJM model is
  related to different subjects such as topological string, see
review in  \volcite{MA}.  On the other hand, the 5d theory establishes a
rather exotic scaling $N^3$
  for the gauge theory, and it supports the relation of the 5d theory to 6d
  $(2,0)$ superconformal field theory, see review in \volcite{KL}.

  Once we start to calculate the partition functions on different
  manifolds (e.g., $S^r$ and $S^{r-1}\times S^1$), we start to realize
the composite structure of the answer. Namely the answer can be built
from  basic objects called holomorphic blocks, this is discussed in
details for 2d, 3d, 4d and 5d theories in \volcite{WI} and \volcite{PA}.
Besides, it seems that in odd dimensions the partition function may
serve as a good measure for the number of degrees of freedom. This can
be made more precise for the partition function on $S^3$ which measures
the number of degrees of freedom of the supersymmetric theory.  Thus one
can study how it behaves along the RG flow, see \volcite{PU}.

Another interesting application of localization appears in the context
of the BPS/CFT-correspondence \cite{PZNekrasov:2004}, in which BPS
phenomena of 4d gauge theories are related to 2d conformal field theory
or its massive, lattice, or integrable deformation.  A beautiful and
precise realization of this
idea is the Alday-Gaiotto-Tachikawa (AGT) correspondence which relates  4d
$\mathcal{N}=2$ gauge theory of class $\mathcal{S}$ to Liouville (Toda) CFT on
some Riemann surface $C$. 
A 4d $\mathcal{N}=2$ gauge theory of class $\mathcal{S}$ is obtained by
compactification of 6d $(2,0)$ tensor self-dual theory on $C$. For a
review of this topic see \volcite{TA}.

The 3d/3d version of this correspondence is
reviewed in  \volcite{DI} and  5d version is reviewed in \volcite{PA}.
        
The 2d supersymmetric non-linear sigma models play a prominent role in
string theory and mathematical physics, but it is hard to perform 
direct calculations for non-linear sigma model.  However some gauged linear sigma
models (2d supersymmetric gauge theories) flow to non-linear sigma
model. This flow allows to compute some quantities of non-linear sigma
models, such as genus 0
Gromov-Witten invariants (counting of holomorphic maps from $S^2 \simeq \mathbb{CP}^1$ to
a Calabi-Yau target) by localization in 2d gauge theories on $S^2$. See review in \volcite{MO}
and \volcite{BL}.
 
Other important applications of localization calculations are explicit
checks of QFT dualities. Sometimes QFT theories with different
Lagrangians describe the same physical system and have the same physical
dynamics, a famous example is Seiberg duality \cite{PZSeiberg:1994pq}. 
The dual theories may look very different in the description by 
gauge group and matter content, but have the same partition functions, provided approriate identification of the parameters. 
Various checks of the duality using the localization results are
reviewed in \volcite{BL}, \volcite{WI}, \volcite{PU} and \volcite{RR}.

{\bf Acknowledgement:}   We thank all authors who contributed to this
volume: Francesco Benini, Tudor Dimofte, Thomas T. Dumitrescu, 
Kazuo Hosomichi, Seok Kim, Kimyeong Lee, 
Bruno Le Floch, 
Marcos Mari\~{n}o, Joseph A. Minahan, David Morrison, 
Sara Pasquetti, Jian Qiu, Leonardo Rastelli, Shlomo S. Razamat, 
 Silvu S. Pufu, Yuji Tachikawa and Brian Willett. 
 We are grateful for their cooperation and for their patience with
long completion of this project, for many suggestions and
 helpful ideas.
Special thanks to Joseph A.~Minahan and Bruno Le Floch for a meticulous 
proofreading of this introduction, and to Guido Festuccia for comments. 

 The research of V.P. is 
supported  by ERC grant QUASIFT. 
The research of M.Z. is supported  by Vetenskapsr\r{a}det under grant \#2014- 5517, by the STINT grant and by the grant 
``Geometry and Physics" from the Knut and Alice Wallenberg foundation.

\documentfinish


%% file: header.tex
\newcommand{\chapterauthor}[1]{
\begin{center}
{\bf \normalsize  #1}
\end{center}
}

\newcommand{\chapteraddress}[1]{
\begin{center}
{ \small \it \addressvalue}
\end{center}
}

\newcommand{\chapterabstract}[1]{
\vspace{\baselineskip}
\begin{center}
\textbf{\small Abstract}
\end{center}
#1}

\newcommand{\chapterheader}{

\chapter[\titlevalue{}  (by \shortauthorvalue)]{\titlevalue}
\label{Chapter\IDvalue}
\chapterauthor{\authorvalue}
\chapteraddress{\addressvalue}
\chapterabstract{\abstractvalue}
\tightmtctrue
\minitoc
}

\newcommand{\documentheader}{
\begin{flushright} \small
  \preprintvalue
 \end{flushright}

\begin{center}
{\bf \Large \titlevalue}
\end{center}

\chapterauthor{\authorvalue}
\chapteraddress{\addressvalue}
\chapterabstract{\abstractvalue}

\medskip

This is a contribution to the review volume ``Localization techniques
in quantum field theories'' (eds. V.~Pestun and M.~Zabzine) which
contains 17 Chapters available at \cite{ContributionSummary}

\tableofcontents
}

\newcommand{\ifvolume}[2]{\ifx\ifLONG\undefined#2\else#1\fi}

\newcommand{\documentfinish}{
\ifx\ifLONG\undefined
\bibliographystyle{bibreview} 
\bibliography{\IDvalue,review}  
\end{document}
\else
\addcontentsline{toc}{section}{References}
\input{\DIRvalue/\IDvalue.bbl}
\fi
}

\newcommand{\documentfinishBBL}{
\addcontentsline{toc}{section}{References}
\ifx\ifLONG\undefined
\input{\IDvalue.separate.bbl}
\end{document}
\else
\input{\DIRvalue/\IDvalue.volume.bbl}
\fi
}

\ifx\ifLONG\undefined
\def\volcite#1{Contribution \cite{Contribution#1}}
\else 
\def\volcite#1{Chapter \ref{Chapter#1}}
\fi

%% file: PZ.bbl
\providecommand{\href}[2]{#2}\begingroup\raggedright\begin{thebibliography}{10}

\bibitem{ContributionSummary}
V.~Pestun and M.~Zabzine, eds., {\em Localization techniques in quantum field
  theory}, vol.~xx.
\newblock Journal of Physics A, 2016.
\newblock \href{http://arxiv.org/abs/1608.02952}{{\tt 1608.02952}}.
\newblock \url{https://arxiv.org/src/1608.02952/anc/LocQFT.pdf},
  \url{http://pestun.ihes.fr/pages/LocalizationReview/LocQFT.pdf}.

\bibitem{PZMR1501331}
S.~Lefschetz, ``Intersections and transformations of complexes and manifolds,''
  \href{http://dx.doi.org/10.2307/1989171}{{\em Trans. Amer. Math. Soc.} {\bf
  28} (1926) no.~1, 1--49}. \url{http://dx.doi.org/10.2307/1989171}.

\bibitem{PZMR674406}
J.~J. Duistermaat and G.~J. Heckman, ``On the variation in the cohomology of
  the symplectic form of the reduced phase space,''
  \href{http://dx.doi.org/10.1007/BF01399506}{{\em Invent. Math.} {\bf 69}
  (1982) no.~2, 259--268}. \url{http://dx.doi.org/10.1007/BF01399506}.

\bibitem{PZMR685019}
N.~Berline and M.~Vergne, ``Classes caract\'eristiques \'equivariantes.
  {F}ormule de localisation en cohomologie \'equivariante,'' {\em C. R. Acad.
  Sci. Paris S\'er. I Math.} {\bf 295} (1982) no.~9, 539--541.

\bibitem{PZMR721448}
M.~F. Atiyah and R.~Bott, ``The moment map and equivariant cohomology,''
  \href{http://dx.doi.org/10.1016/0040-9383(84)90021-1}{{\em Topology} {\bf 23}
  (1984) no.~1, 1--28}. \url{http://dx.doi.org/10.1016/0040-9383(84)90021-1}.

\bibitem{ContributionPE}
V.~Pestun, ``Review of localization in geometry,'' {\em Journal of Physics A}
  {\bf xx} (2016)  000, \href{http://arxiv.org/abs/1608.02954}{{\tt
  1608.02954}}.

\bibitem{PZMR683171}
E.~Witten, ``Supersymmetry and {M}orse theory,'' {\em J. Differential Geom.}
  {\bf 17} (1982) no.~4, 661--692 (1983).
  \url{http://projecteuclid.org/euclid.jdg/1214437492}.

\bibitem{PZMR958805}
E.~Witten, ``Topological sigma models,'' {\em Comm. Math. Phys.} {\bf 118}
  (1988) no.~3, 411--449. \url{http://projecteuclid.org/euclid.cmp/1104162092}.

\bibitem{PZMR953828}
E.~Witten, ``Topological quantum field theory,'' {\em Comm. Math. Phys.} {\bf
  117} (1988) no.~3, 353--386.
  \url{http://projecteuclid.org/euclid.cmp/1104161738}.

\bibitem{PZWitten:1992xu}
E.~Witten, ``{Two-dimensional gauge theories revisited},''
  \href{http://dx.doi.org/10.1016/0393-0440(92)90034-X}{{\em J. Geom. Phys.}
  {\bf 9} (1992)  303--368},
\href{http://arxiv.org/abs/hep-th/9204083}{{\tt arXiv:hep-th/9204083
  [hep-th]}}.

\bibitem{PZMR2045303}
N.~A. Nekrasov, ``Seiberg-{W}itten prepotential from instanton counting,'' {\em
  Adv. Theor. Math. Phys.} {\bf 7} (2003) no.~5, 831--864.
  \url{http://projecteuclid.org/euclid.atmp/1111510432}.

\bibitem{PZLosev:1997tp}
A.~Losev, N.~Nekrasov, and S.~L. Shatashvili, ``{Issues in topological gauge
  theory},'' \href{http://dx.doi.org/10.1016/S0550-3213(98)00628-2}{{\em Nucl.
  Phys.} {\bf B534} (1998)  549--611},
\href{http://arxiv.org/abs/hep-th/9711108}{{\tt arXiv:hep-th/9711108
  [hep-th]}}.

\bibitem{PZMoore:1997dj}
G.~W. Moore, N.~Nekrasov, and S.~Shatashvili, ``{Integrating over Higgs
  branches},'' \href{http://dx.doi.org/10.1007/PL00005525}{{\em Commun. Math.
  Phys.} {\bf 209} (2000)  97--121},
\href{http://arxiv.org/abs/hep-th/9712241}{{\tt arXiv:hep-th/9712241
  [hep-th]}}.

\bibitem{PZLossev:1997bz}
A.~Lossev, N.~Nekrasov, and S.~L. Shatashvili, ``{Testing Seiberg-Witten
  solution},'' \href{http://arxiv.org/abs/hep-th/9801061}{{\tt
  arXiv:hep-th/9801061 [hep-th]}}.
\url{http://alice.cern.ch/format/showfull?sysnb=0266564}.

\bibitem{PZLosev:1995cr}
A.~Losev, G.~W. Moore, N.~Nekrasov, and S.~Shatashvili, ``{Four-dimensional
  avatars of two-dimensional RCFT},''
  \href{http://dx.doi.org/10.1016/0920-5632(96)00015-1}{{\em Nucl. Phys. Proc.
  Suppl.} {\bf 46} (1996)  130--145},
\href{http://arxiv.org/abs/hep-th/9509151}{{\tt arXiv:hep-th/9509151
  [hep-th]}}.

\bibitem{PZPestun:2007rz}
V.~Pestun, ``{Localization of gauge theory on a four-sphere and supersymmetric
  Wilson loops},'' \href{http://dx.doi.org/10.1007/s00220-012-1485-0}{{\em
  Commun.Math.Phys.} {\bf 313} (2012)  71--129},
\href{http://arxiv.org/abs/0712.2824}{{\tt arXiv:0712.2824 [hep-th]}}.

\bibitem{PZBlau:2000xg}
M.~Blau, ``{Killing spinors and SYM on curved spaces},''
  \href{http://dx.doi.org/10.1088/1126-6708/2000/11/023}{{\em JHEP} {\bf 11}
  (2000)  023},
\href{http://arxiv.org/abs/hep-th/0005098}{{\tt arXiv:hep-th/0005098
  [hep-th]}}.

\bibitem{PZFestuccia:2011ws}
G.~Festuccia and N.~Seiberg, ``{Rigid Supersymmetric Theories in Curved
  Superspace},'' \href{http://dx.doi.org/10.1007/JHEP06(2011)114}{{\em JHEP}
  {\bf 06} (2011)  114},
\href{http://arxiv.org/abs/1105.0689}{{\tt arXiv:1105.0689 [hep-th]}}.

\bibitem{PZImbimbo:2014pla}
C.~Imbimbo and D.~Rosa, ``{Topological anomalies for Seifert 3-manifolds},''
  \href{http://dx.doi.org/10.1007/JHEP07(2015)068}{{\em JHEP} {\bf 07} (2015)
  068},
\href{http://arxiv.org/abs/1411.6635}{{\tt arXiv:1411.6635 [hep-th]}}.

\bibitem{PZBae:2015eoa}
J.~Bae, C.~Imbimbo, S.-J. Rey, and D.~Rosa, ``{New Supersymmetric Localizations
  from Topological Gravity},''
  \href{http://dx.doi.org/10.1007/JHEP03(2016)169}{{\em JHEP} {\bf 03} (2016)
  169},
\href{http://arxiv.org/abs/1510.00006}{{\tt arXiv:1510.00006 [hep-th]}}.

\bibitem{PZKlare:2012gn}
C.~Klare, A.~Tomasiello, and A.~Zaffaroni, ``{Supersymmetry on Curved Spaces
  and Holography},'' \href{http://dx.doi.org/10.1007/JHEP08(2012)061}{{\em
  JHEP} {\bf 08} (2012)  061},
\href{http://arxiv.org/abs/1205.1062}{{\tt arXiv:1205.1062 [hep-th]}}.

\bibitem{PZDumitrescu:2012ha}
T.~T. Dumitrescu, G.~Festuccia, and N.~Seiberg, ``{Exploring Curved
  Superspace},'' \href{http://dx.doi.org/10.1007/JHEP08(2012)141}{{\em JHEP}
  {\bf 08} (2012)  141},
\href{http://arxiv.org/abs/1205.1115}{{\tt arXiv:1205.1115 [hep-th]}}.

\bibitem{PZCassani:2012ri}
D.~Cassani, C.~Klare, D.~Martelli, A.~Tomasiello, and A.~Zaffaroni,
  ``{Supersymmetry in Lorentzian Curved Spaces and Holography},''
  \href{http://dx.doi.org/10.1007/s00220-014-1983-3}{{\em Commun. Math. Phys.}
  {\bf 327} (2014)  577--602},
\href{http://arxiv.org/abs/1207.2181}{{\tt arXiv:1207.2181 [hep-th]}}.

\bibitem{PZClosset:2012ru}
C.~Closset, T.~T. Dumitrescu, G.~Festuccia, and Z.~Komargodski,
  ``{Supersymmetric Field Theories on Three-Manifolds},''
  \href{http://dx.doi.org/10.1007/JHEP05(2013)017}{{\em JHEP} {\bf 05} (2013)
  017},
\href{http://arxiv.org/abs/1212.3388}{{\tt arXiv:1212.3388 [hep-th]}}.

\bibitem{PZKlare:2013dka}
C.~Klare and A.~Zaffaroni, ``{Extended Supersymmetry on Curved Spaces},''
  \href{http://dx.doi.org/10.1007/JHEP10(2013)218}{{\em JHEP} {\bf 10} (2013)
  218},
\href{http://arxiv.org/abs/1308.1102}{{\tt arXiv:1308.1102 [hep-th]}}.

\bibitem{PZButter:2015tra}
D.~Butter, G.~Inverso, and I.~Lodato, ``{Rigid 4D $ \mathcal{N}=2 $
  supersymmetric backgrounds and actions},''
  \href{http://dx.doi.org/10.1007/JHEP09(2015)088}{{\em JHEP} {\bf 09} (2015)
  088},
\href{http://arxiv.org/abs/1505.03500}{{\tt arXiv:1505.03500 [hep-th]}}.

\bibitem{PZPestun:2014mja}
V.~Pestun, \href{http://dx.doi.org/10.1007/978-3-319-18769-3_6}{``{Localization
  for $\mathcal {N}=2$ Supersymmetric Gauge Theories in Four Dimensions},''} in
  {\em New Dualities of Supersymmetric Gauge Theories}, J.~Teschner, ed.,
  pp.~159--194.
\newblock 2016.
\newblock \href{http://arxiv.org/abs/1412.7134}{{\tt arXiv:1412.7134
  [hep-th]}}.
\newblock
\url{http://inspirehep.net/record/1335346/files/arXiv:1412.7134.pdf}.
\newblock

\bibitem{PZPan:2013uoa}
Y.~Pan, ``{Rigid Supersymmetry on 5-dimensional Riemannian Manifolds and
  Contact Geometry},'' \href{http://dx.doi.org/10.1007/JHEP05(2014)041}{{\em
  JHEP} {\bf 05} (2014)  041},
\href{http://arxiv.org/abs/1308.1567}{{\tt arXiv:1308.1567 [hep-th]}}.

\bibitem{PZImamura:2014ima}
Y.~Imamura and H.~Matsuno, ``{Supersymmetric backgrounds from 5d N=1
  supergravity},'' \href{http://dx.doi.org/10.1007/JHEP07(2014)055}{{\em JHEP}
  {\bf 07} (2014)  055},
\href{http://arxiv.org/abs/1404.0210}{{\tt arXiv:1404.0210 [hep-th]}}.

\bibitem{PZPan:2015nba}
Y.~Pan and J.~Schmude, ``{On rigid supersymmetry and notions of holomorphy in
  five dimensions},'' \href{http://dx.doi.org/10.1007/JHEP11(2015)041}{{\em
  JHEP} {\bf 11} (2015)  041},
\href{http://arxiv.org/abs/1504.00321}{{\tt arXiv:1504.00321 [hep-th]}}.

\bibitem{PZSamtleben:2012ua}
H.~Samtleben, E.~Sezgin, and D.~Tsimpis, ``{Rigid 6D supersymmetry and
  localization},'' \href{http://dx.doi.org/10.1007/JHEP03(2013)137}{{\em JHEP}
  {\bf 03} (2013)  137},
\href{http://arxiv.org/abs/1212.4706}{{\tt arXiv:1212.4706 [hep-th]}}.

\bibitem{PZKuzenko:2015lca}
S.~M. Kuzenko, ``{Supersymmetric Spacetimes from Curved Superspace},'' {\em
  PoS} {\bf CORFU2014} (2015)  140,
\href{http://arxiv.org/abs/1504.08114}{{\tt arXiv:1504.08114 [hep-th]}}.

\bibitem{PZKuzenko:2014eqa}
S.~M. Kuzenko, J.~Novak, and G.~Tartaglino-Mazzucchelli, ``{Symmetries of
  curved superspace in five dimensions},''
  \href{http://dx.doi.org/10.1007/JHEP10(2014)175}{{\em JHEP} {\bf 10} (2014)
  175},
\href{http://arxiv.org/abs/1406.0727}{{\tt arXiv:1406.0727 [hep-th]}}.

\bibitem{ContributionDU}
T.~Dumitrescu, ``An Introduction to Supersymmetric Field Theories in Curved
  Space,'' {\em Journal of Physics A} {\bf xx} (2016)  000,
  \href{http://arxiv.org/abs/1608.02957}{{\tt 1608.02957}}.

\bibitem{PZKapustin:2009kz}
A.~Kapustin, B.~Willett, and I.~Yaakov, ``{Exact Results for Wilson Loops in
  Superconformal Chern-Simons Theories with Matter},''
  \href{http://dx.doi.org/10.1007/JHEP03(2010)089}{{\em JHEP} {\bf 03} (2010)
  089},
\href{http://arxiv.org/abs/0909.4559}{{\tt arXiv:0909.4559 [hep-th]}}.

\bibitem{PZBenini:2012ui}
F.~Benini and S.~Cremonesi, ``{Partition Functions of ${\mathcal{N}=(2,2)}$
  Gauge Theories on S$^{2}$ and Vortices},''
  \href{http://dx.doi.org/10.1007/s00220-014-2112-z}{{\em Commun. Math. Phys.}
  {\bf 334} (2015) no.~3, 1483--1527},
\href{http://arxiv.org/abs/1206.2356}{{\tt arXiv:1206.2356 [hep-th]}}.

\bibitem{PZDoroud:2012xw}
N.~Doroud, J.~Gomis, B.~Le~Floch, and S.~Lee, ``{Exact Results in D=2
  Supersymmetric Gauge Theories},''
  \href{http://dx.doi.org/10.1007/JHEP05(2013)093}{{\em JHEP} {\bf 05} (2013)
  093},
\href{http://arxiv.org/abs/1206.2606}{{\tt arXiv:1206.2606 [hep-th]}}.

\bibitem{PZKallen:2012cs}
J.~Kallen and M.~Zabzine, ``{Twisted supersymmetric 5D Yang-Mills theory and
  contact geometry},'' \href{http://dx.doi.org/10.1007/JHEP05(2012)125}{{\em
  JHEP} {\bf 05} (2012)  125},
\href{http://arxiv.org/abs/1202.1956}{{\tt arXiv:1202.1956 [hep-th]}}.

\bibitem{PZKallen:2012va}
J.~Kallen, J.~Qiu, and M.~Zabzine, ``{The perturbative partition function of
  supersymmetric 5D Yang-Mills theory with matter on the five-sphere},''
  \href{http://dx.doi.org/10.1007/JHEP08(2012)157}{{\em JHEP} {\bf 08} (2012)
  157},
\href{http://arxiv.org/abs/1206.6008}{{\tt arXiv:1206.6008 [hep-th]}}.

\bibitem{PZKim:2012ava}
H.-C. Kim and S.~Kim, ``{M5-branes from gauge theories on the 5-sphere},''
  \href{http://dx.doi.org/10.1007/JHEP05(2013)144}{{\em JHEP} {\bf 05} (2013)
  144},
\href{http://arxiv.org/abs/1206.6339}{{\tt arXiv:1206.6339 [hep-th]}}.

\bibitem{PZMinahan:2015jta}
J.~A. Minahan and M.~Zabzine, ``{Gauge theories with 16 supersymmetries on
  spheres},'' \href{http://dx.doi.org/10.1007/JHEP03(2015)155}{{\em JHEP} {\bf
  03} (2015)  155},
\href{http://arxiv.org/abs/1502.07154}{{\tt arXiv:1502.07154 [hep-th]}}.

\bibitem{PZHama:2011ea}
N.~Hama, K.~Hosomichi, and S.~Lee, ``{SUSY Gauge Theories on Squashed
  Three-Spheres},'' \href{http://dx.doi.org/10.1007/JHEP05(2011)014}{{\em JHEP}
  {\bf 05} (2011)  014},
\href{http://arxiv.org/abs/1102.4716}{{\tt arXiv:1102.4716 [hep-th]}}.

\bibitem{PZImamura:2011wg}
Y.~Imamura and D.~Yokoyama, ``{N=2 supersymmetric theories on squashed
  three-sphere},'' \href{http://dx.doi.org/10.1103/PhysRevD.85.025015}{{\em
  Phys. Rev.} {\bf D85} (2012)  025015},
\href{http://arxiv.org/abs/1109.4734}{{\tt arXiv:1109.4734 [hep-th]}}.

\bibitem{PZHama:2012bg}
N.~Hama and K.~Hosomichi, ``{Seiberg-Witten Theories on Ellipsoids},''
  \href{http://dx.doi.org/10.1007/JHEP09(2012)033,
  10.1007/JHEP10(2012)051}{{\em JHEP} {\bf 09} (2012)  033},
  \href{http://arxiv.org/abs/1206.6359}{{\tt arXiv:1206.6359 [hep-th]}}.
[Addendum: JHEP10,051(2012)].

\bibitem{PZLockhart:2012vp}
G.~Lockhart and C.~Vafa, ``{Superconformal Partition Functions and
  Non-perturbative Topological Strings},''
\href{http://arxiv.org/abs/1210.5909}{{\tt arXiv:1210.5909 [hep-th]}}.

\bibitem{PZImamura:2012bm}
Y.~Imamura, ``{Perturbative partition function for squashed $S^5$},''
\href{http://arxiv.org/abs/1210.6308}{{\tt arXiv:1210.6308 [hep-th]}}.

\bibitem{PZKim:2012qf}
H.-C. Kim, J.~Kim, and S.~Kim, ``{Instantons on the 5-sphere and M5-branes},''
\href{http://arxiv.org/abs/1211.0144}{{\tt arXiv:1211.0144 [hep-th]}}.

\bibitem{PZMinahan:2015any}
J.~A. Minahan, ``{Localizing gauge theories on S$^{d}$},''
  \href{http://dx.doi.org/10.1007/JHEP04(2016)152}{{\em JHEP} {\bf 04} (2016)
  152},
\href{http://arxiv.org/abs/1512.06924}{{\tt arXiv:1512.06924 [hep-th]}}.

\bibitem{PZNarukawa2004247}
A.~Narukawa, ``The modular properties and the integral representations of the
  multiple elliptic gamma functions,''
  \href{http://dx.doi.org/http://dx.doi.org/10.1016/j.aim.2003.11.009}{{\em
  Advances in Mathematics} {\bf 189} (2004) no.~2, 247 -- 267}.
  \url{http://www.sciencedirect.com/science/article/pii/S0001870803003670}.

\bibitem{ContributionBL}
F.~Benini and B.~{Le Floch}, ``Supersymmetric localization in two dimensions,''
  {\em Journal of Physics A} {\bf xx} (2016)  000,
  \href{http://arxiv.org/abs/1608.02955}{{\tt 1608.02955}}.

\bibitem{ContributionHO}
K.~Hosomichi, ``$\mathcal{N}=2$ SUSY gauge theories on $S^4$,'' {\em Journal of
  Physics A} {\bf xx} (2016)  000, \href{http://arxiv.org/abs/1608.02962}{{\tt
  1608.02962}}.

\bibitem{ContributionWI}
B.~Willett, ``Localization on three-dimensional manifolds,'' {\em Journal of
  Physics A} {\bf xx} (2016)  000, \href{http://arxiv.org/abs/1608.02958}{{\tt
  1608.02958}}.

\bibitem{ContributionQZ}
J.~Qiu and M.~Zabzine, ``Review of localization for 5D supersymmetric gauge
  theories,'' {\em Journal of Physics A} {\bf xx} (2016)  000,
  \href{http://arxiv.org/abs/1608.02966}{{\tt 1608.02966}}.

\bibitem{PZAtiyah1974}
M.~F. Atiyah, {\em Elliptic operators and compact groups}.
\newblock Springer-Verlag, Berlin, 1974.
\newblock Lecture Notes in Mathematics, Vol. 401.

\bibitem{PZZamolodchikov:1995aa}
A.~B. Zamolodchikov and A.~B. Zamolodchikov, ``{Structure constants and
  conformal bootstrap in Liouville field theory},''
  \href{http://dx.doi.org/10.1016/0550-3213(96)00351-3}{{\em Nucl. Phys.} {\bf
  B477} (1996)  577--605},
\href{http://arxiv.org/abs/hep-th/9506136}{{\tt arXiv:hep-th/9506136
  [hep-th]}}.

\bibitem{PZAlday:2009aq}
L.~F. Alday, D.~Gaiotto, and Y.~Tachikawa, ``{Liouville Correlation Functions
  from Four-dimensional Gauge Theories},''
  \href{http://dx.doi.org/10.1007/s11005-010-0369-5}{{\em Lett.Math.Phys.} {\bf
  91} (2010)  167--197},
\href{http://arxiv.org/abs/0906.3219}{{\tt arXiv:0906.3219 [hep-th]}}.

\bibitem{ContributionTA}
Y.~Tachikawa, ``A brief review of the 2d/4d correspondences,'' {\em Journal of
  Physics A} {\bf xx} (2016)  000, \href{http://arxiv.org/abs/1608.02964}{{\tt
  1608.02964}}.

\bibitem{ContributionMA}
M.~Mari{\~n}o, ``Localization at large $N$ in Chern-Simons-matter theories,''
  {\em Journal of Physics A} {\bf xx} (2016)  000,
  \href{http://arxiv.org/abs/1608.02959}{{\tt 1608.02959}}.

\bibitem{ContributionPU}
S.~Pufu, ``The F-Theorem and F-Maximization,'' {\em Journal of Physics A} {\bf
  xx} (2016)  000, \href{http://arxiv.org/abs/1608.02960}{{\tt 1608.02960}}.

\bibitem{ContributionDI}
T.~Dimofte, ``Perturbative and nonperturbative aspects of complex Chern-Simons
  Theory,'' {\em Journal of Physics A} {\bf xx} (2016)  000,
  \href{http://arxiv.org/abs/1608.02961}{{\tt 1608.02961}}.

\bibitem{ContributionMI}
J.~Minahan, ``Matrix models for 5D super Yang-Mills,'' {\em Journal of Physics
  A} {\bf xx} (2016)  000, \href{http://arxiv.org/abs/1608.02967}{{\tt
  1608.02967}}.

\bibitem{ContributionPA}
S.~Pasquetti, ``Holomorphic blocks and the 5d AGT correspondence,'' {\em
  Journal of Physics A} {\bf xx} (2016)  000,
  \href{http://arxiv.org/abs/1608.02968}{{\tt 1608.02968}}.

\bibitem{ContributionRR}
L.~Rastelli and S.~Razamat, ``The supersymmetric index in four dimensions,''
  {\em Journal of Physics A} {\bf xx} (2016)  000,
  \href{http://arxiv.org/abs/1608.02965}{{\tt 1608.02965}}.

\bibitem{ContributionKL}
S.~Kim and K.~Lee, ``Indices for 6 dimensional superconformal field theories,''
  {\em Journal of Physics A} {\bf xx} (2016)  000,
  \href{http://arxiv.org/abs/1608.02969}{{\tt 1608.02969}}.

\bibitem{ContributionZA}
K.~Zarembo, ``Localization and AdS/CFT Correspondence,'' {\em Journal of
  Physics A} {\bf xx} (2016)  000, \href{http://arxiv.org/abs/1608.02963}{{\tt
  1608.02963}}.

\bibitem{PZNekrasov:2004}
N.~Nekrasov, ``{On the BPS/CFT correspondence},'' {\em Lecture at the
  University of Amsterdam string theory group seminar} (2004)  .

\bibitem{ContributionMO}
D.~Morrison, ``Gromov-Witten invariants and localization,'' {\em Journal of
  Physics A} (2016)  , \href{http://arxiv.org/abs/1608.02956}{{\tt
  1608.02956}}.

\bibitem{PZSeiberg:1994pq}
N.~Seiberg, ``{Electric - magnetic duality in supersymmetric nonAbelian gauge
  theories},'' \href{http://dx.doi.org/10.1016/0550-3213(94)00023-8}{{\em Nucl.
  Phys.} {\bf B435} (1995)  129--146},
\href{http://arxiv.org/abs/hep-th/9411149}{{\tt arXiv:hep-th/9411149
  [hep-th]}}.

\end{thebibliography}\endgroup

%% file: PEdef.tex
\usepackage{ytableau}
\usepackage[margin=1in]{geometry}
\usepackage{amsmath,amssymb,amsfonts,mathtools}
\usepackage{amsxtra}
\usepackage{amscd}
\usepackage{xcolor}

 \usepackage{tikz-cd}
 \usetikzlibrary{matrix,arrows,decorations.pathmorphing}
 \usepackage[mathscr]{euscript}
 \usepackage{amscd}
 \usepackage{slashed}
 \usepackage{listings}
 \usepackage{graphicx}
 \usepackage{todonotes}

\usepackage[pagebackref=false]{hyperref}
 \definecolor{darkblue}{RGB}{0,0,192}
 \definecolor{darkgreen}{RGB}{0,192,0}
\hypersetup{colorlinks = true,extension = notused,linkcolor = darkblue,anchorcolor = red,citecolor = darkgreen,filecolor = red,pagecolor = red,urlcolor = darkblue}

 \usepackage{iftex}
 \ifxetex
         \usepackage{fontspec}
 \else
 \fi

\newcommand {\BR}   {\mathbb R}

\newcommand {\BC}   {\mathbb C}

\newcommand {\BO}   {\mathbb O}

\newcommand {\CP}   {\mathbb C \mathbb P}

 \newcommand {\ep}  {\epsilon}

\newcommand{\g}{\mathfrak{g}}

\DeclareMathOperator{\ch}{ch}

\DeclareMathOperator{\Tr} {Tr}

\DeclareMathOperator{\ind} {ind}

\newcommand{\SU}{SU}

\numberwithin{equation}{section}